\documentclass[12pt]{iopart}

\usepackage{graphicx}

\usepackage[dvipdfm]{hyperref}

\usepackage{amssymb}

\usepackage{cite}

\usepackage[final]{changes}

\newcommand{\subtxt}[1]{\mbox{\scriptsize #1}}

\newcommand{\PoP}[0]{Phys. Plasmas}
\newcommand{\NF}[0]{Nucl. Fusion}
\newcommand{\JPP}[0]{J. Plasma Phys.}
\newcommand{\CPC}[0]{Comp. Phys. Comm.}
\newcommand{\FST}[0]{Fusion Science and Technology}

\begin{document}

\title[Microwave beam broadening due to $\tilde{n}_e$ ]{Microwave beam broadening due to turbulent plasma density fluctuations within the limit of the Born approximation and beyond}

\author{A~K\"{o}hn$^{1,2}$, L~Guidi$^{1,3}$, E~Holzhauer$^2$, O~Maj$^1$, E~Poli$^1$, A~Snicker$^{1,4}$, H~Weber$^{1}$}

\address{$^1$Max Planck Institute for Plasma Physics, D-85748 Garching, Germany}
\address{$^2$Institute of Interfacial Process Engineering and Plasma Technology, University of Stuttgart, D-70569 Stuttgart, Germany}
\address{$^3$Technische Universit\"{a}t M\"{u}nchen, Numerical Methods for Plasma Physics (M16), D-85748 Garching, Germany}
\address{$^4$Department of Applied Physics, Aalto University, FI- 00076 Aalto, Finland}

\ead{koehn@igvp.uni-stuttgart.de}
\begin{abstract}
Plasma turbulence, and edge density fluctuations in particular, can under certain conditions broaden the cross-section of injected microwave beams significantly. This can be a severe problem for applications relying on well-localized deposition of the microwave power, like the control of MHD instabilities. Here we investigate this broadening mechanism as a function of fluctuation level, background density and propagation length in a fusion-relevant scenario using two numerical codes, the full-wave code IPF-FDMC and the novel wave kinetic equation solver WKBeam. The latter treats the effects of fluctuations using a statistical approach, based on an iterative solution of the scattering problem (Born approximation). The full-wave simulations are used to benchmark this approach.
The Born approximation is shown to be valid over a large parameter range, including ITER-relevant scenarios.
\end{abstract}

\submitto{\PPCF}

\section{Introduction}\label{s:intro}
Electromagnetic waves in the microwave range of frequencies are widely used in fusion-relevant experiments for heating and diagnostic purposes~\cite{Hartfuss.2014,Bornatici.1983,Prater.2004}. In tokamaks, they are employed among others for control and suppression of MHD instabilities like the sawtooth oscillation and the neoclassical tearing mode (NTM)~\cite{Igochine.2015}. These applications require a good localization of the deposited wave power~\cite{Westerhof.1987}. In particular, NTMs can lead to a degradation of the confinement up to a disruption of the discharge~\cite{LaHaye.2006}. Since NTMs are driven by small perturbations in the plasma current profile (more precisely in the bootstrap current profile) resulting in the formation of magnetic islands, one way to stabilize them is to drive currents at the islands' positions and restore the original current profile. This can be achieved by injecting microwaves in the electron cyclotron range of frequencies~\cite{Zohm.2007,Kasparek.2016}. It requires however a precise spatial localization of the place of absorption as the current should be ideally driven within the islands~\cite{Sauter.2010}.

\replaced{To}{The} reach the magnetic surface on which the NTM develops, the injected microwave beam has to pass the plasma boundary where density fluctuations occur with amplitudes up to $100\,\%$~\cite{Zweben.2007}. This can significantly distort the beam and thus spoil the good localization, strongly reducing the efficiency of the NTM stabilization. A sound understanding of this effect is mandatory in order to predict the effectiveness of the microwave beams for the control tasks described above. \deleted{This paper contributes to that understanding with the aid of numerical simulations. Two different codes are used: the full-wave code IPF-FDMC~\mbox{\cite{Koehn.2008}} and the WKBeam code~\mbox{\cite{Weber.2015}} which solves the wave kinetic equation in the presence of random fluctuations in the background density. While for the first code an ensemble of different realizations of turbulent density fluctuations is required to reproduce the situation in the experiment, WKBeam allows to directly calculate the average effect by applying a statistical operator. The derivation of this scattering operator is based on the so-called Born approximation~\mbox{\cite{Born-Wolf.1999}}. The WKBeam results are thus expected to become invalid for sufficiently high fluctuation levels. These limitations of the latter treatment are explored and quantified in this paper.} 

The influence of edge plasma density fluctuations on injected microwaves has been studied with geometrical-optics tools in the 1980s in a fusion-relevant context when high-power microwave sources became available~\cite{Ott.1980,Hui.1981,Hansen.1988}. The topic has been brought back into focus by Tsironis in 2009~\cite{Tsironis.2009} which triggered a significant follow-up research looking into this problem using different techniques~\cite{Bertelli.2010,Peysson.2011,Balakin.2011,Ram.2013,Ram.2016,Poli.2015,Sysoeva.2015,Ioannidis.2017}. As a common agreement one can state that \emph{(a)} substantial broadening of microwave beams due to edge plasma density perturbations is expected, \emph{(b)} the situation in medium-sized tokamaks differs from large-scale tokamaks like ITER (due to differences in microwave frequency, size of turbulent structures, and propagation length), \emph{(c)} further and more detailed studies with a minimum of simplifying assumptions are needed for the ITER scenarios which cannot be explored experimentally in today's tokamaks, and \emph{(d)} the various numerical tools should be cross-benchmarked.

\added{This paper contributes to the understanding of the interaction of microwaves with turbulent plasma density fluctuations with the aid of numerical simulations. Two different codes are used: the full-wave code IPF-FDMC~\cite{Koehn.2008} and the WKBeam code~\cite{Weber.2015} which solves the wave kinetic equation in the presence of random fluctuations in the background density. While for the first code an ensemble of different realizations of turbulent density fluctuations is required to reproduce the situation in the experiment, WKBeam allows to directly calculate the average effect by applying a statistical operator. The derivation of this scattering operator is based on the so-called Born approximation~\cite{Born-Wolf.1999}. The WKBeam results are thus expected to become invalid for sufficiently high fluctuation levels. These limitations of the latter treatment are explored and quantified.}

This paper is the continuation of previous full-wave simulations of scattering from singular blob-like density structures~\cite{Williams.2014} and of first simulations including turbulent electron density fluctuations~\cite{Koehn.2016}. It serves as a benchmark for the WKBeam code and the recently published results~\cite{Snicker.2018,Snicker.2018B}. The paper is organized as follows: both numerical codes and the generation of the electron density fluctuations are described in Sec.~\ref{s:numerics}. Section~\ref{s:setup} describes the setup of the simulations, followed by Sec.~\ref{s:analysis} which explains how the data obtained from both types of simulations is analyzed. In Sec.~\ref{s:A0_scan_Omode}, the influence of the level of the electron density fluctuations on beam broadening is investigated. The role of the absolute value of the background density is then discussed in Sec.~\ref{s:shift_scan} and the influence of the thickness of the fluctuation layer is discussed in Sec.~\ref{s:wfluctScan}. Results for changing the injected mode from ordinary (O) to the extra-ordinary (X) are presented in Sec.~\ref{s:O_X_mode}. The summary in Sec.~\ref{s:summary} concludes the paper.

\section{Numerics}\label{s:numerics}
This section describes the numerical tools which are used throughout the paper. First, the full-wave code is introduced in Sec.~\ref{s:ipf-fdmc}, followed by the WKBeam code in Sec.~\ref{s:WKBeam}. Both codes are only briefly described and the interested reader is referred to the references given in the corresponding sections. \added{Note that a cold plasma model is used here. The most dangerous NTMs in ITER are expected to occur at radial positions corresponding to electron temperatures of approximately $7\,\mathrm{keV}$~\cite{Snicker.2018}. The effective refractive index changes only marginally for these temperature, see e.g.\ Ref.~\cite{Hartfuss.2014}, and the corresponding effects on the microwave beam propagation are negligible compared to the effect of density fluctuations.}\deleted{ Finally, the generation of synthetic density fluctuations is outlined.}

\subsection{The full-wave code IPF-FDMC}\label{s:ipf-fdmc}
IPF-FDMC~\cite{Koehn.2008} is a finite-difference time-domain (FDTD)~\cite{Taflove.2000} code solving Maxwell's equations and the fluid equation of motion of the electrons on a 2D Cartesian grid. It allows to simulate propagation of electromagnetic waves in a cold magnetized plasma. Specifically, the mathematical model considered consists in evolution equations for the magnetic field $\mathbf{B}$, the electric field $\mathbf{E}$ and the current density $\mathbf{J}$ of the wave, in a plasma equilibrium with background magnetic field $\mathbf{B}_0$ and electron density $n_e$. Specifically,
\begin{eqnarray}
	\frac{\partial}{\partial t}\mathbf{B} &=& -\nabla\times\mathbf{E}\label{e:maxwell1}\\
	\frac{\partial}{\partial t}\mathbf{E} &=& c^2\nabla\times\mathbf{B}-\mathbf{J}/\epsilon_0 \label{e:maxwell2}\\
	\frac{\partial}{\partial t}\mathbf{J} &=& \epsilon_0\omega_{pe}^2\mathbf{E} 
		- \omega_{ce}\mathbf{J}\times\mathbf{\hat{B}}_0 - \nu_e\mathbf{J}\label{e:ipffdmc_J}
\end{eqnarray}
with $c$ the speed of light, $\omega_{pe}=\sqrt{n_ee^2/(\epsilon_0 m_e)}$ the electron plasma frequency, $\omega_{ce}=|e|B_0/m_e$ the electron cyclotron frequency, and 
$\mathbf{\hat{B}}_0$ the unit vector into the direction of the background magnetic field. An electron collision frequency $\nu_e$ is included in Eq.~(\ref{e:ipffdmc_J}) as a dissipation mechanism. The code has been successfully benchmarked against cold plasma theory~\cite{Koehn.2015} and used to study mode conversion processes~\cite{Koehn.2008} and microwave heating in plasmas~\cite{Koehn.2010}.

For the implementation of Eqs.~(\ref{e:maxwell1})-(\ref{e:ipffdmc_J}), the standard FDTD scheme~\cite{Taflove.2000} has to be complemented by a discretization scheme for the current equation~(\ref{e:ipffdmc_J}). Here we use a ''straight forward'' way, that is first advance $J_x$ using the old values of $J_y$ and $J_z$ (in the cross product with the background magnetic field), then advance $J_y$ using the updated value of $J_x$ and the old value of $J_z$, and finally advance $J_z$ using the updated values of $J_x$ and $J_y$. As has been shown in previous works, this method is completely sufficient for a rather large set of problems~\cite{Koehn.2008,Koehn.2010,Lechte.2009}. More advanced methods exist for situations with extreme density gradients and we would like to refer the interested reader to the detailed and thorough analysis by Heuraux and da Silva~\cite{Heuraux.2014,Heuraux.2015,daSilva.2015}.

\subsection{The WKBeam code and the Born scattering approximation}\label{s:WKBeam}
The WKBeam code~\cite{Weber.2015} is based on the formalism of the wave kinetic equation~\cite{McDonald.1988,McDonald.1991} which describes the \emph{average} effect of plasma density fluctuations on a traversing microwave beam. The derivation of scattering operator in the wave kinetic equation relies on the Born approximation which is expected to hold for a weakly scattering medium in the sense specified below. No restrictions need to be made on the spatial size of the turbulent density structures, where other methods based on the short-wavelength approximation (e.g.\ geometrical optics, beam tracing) fail in presence of short-scale fluctuations. The Born approximation imposes, however, a limitation on the amplitude of the turbulent density fluctuations, to be explored in detail in this paper. Since this point is essential for the following discussion, some details about the derivation of the WKBeam model are reported below. The reader is referred to Refs.~\cite{Weber.2015, Snicker.2018} for more details. 

In WKBeam, turbulence is described as a time-independent random field of density fluctuations, with the idea that the time average of a physical quantity of interests can be computed as the ensemble average over a sufficiently large number of samples of the random field. The wave beam is modeled again by  equations~(\ref{e:maxwell1})-(\ref{e:ipffdmc_J}). With the plasma frequency $\omega_{pe}$ being a time-independent random field, we can Fourier transform in time. In the frequency domain Eqs.~(\ref{e:maxwell1})-(\ref{e:ipffdmc_J}) can be written as a single equation for the Fourier transformed wave electric field $\hat{\mathbf{E}} = \hat{\mathbf{E}} (\omega, \mathbf{x})$, namely,
\begin{equation}
  \label{e:Helmholtz}
  \nabla \times ( \nabla \times \hat{\mathbf{E}} )
  - k_0^2 \Big(\hat{\varepsilon}_0 + \frac{\delta n_e}{n_{e,0}} 
  (\hat{\varepsilon}_0 - I) \Big) \hat{\mathbf{E}} = 0,
\end{equation}
where $\hat{\varepsilon}_0 = \hat{\varepsilon}_0(\omega, \mathbf{x})$ is the cold plasma dielectric tensor \cite{Bornatici.1983} computed with the unperturbed electron density $n_{e,0}$, $I$ is the identity tensor, and $\delta n_e$ the random fluctuation field. We assume that the expectation value is $\langle \delta n_e (\mathbf{x})\rangle = 0$, and the correlation function $\langle \delta n_e(\mathbf{x}) \delta n_e(\mathbf{x}') \rangle$ is known. The Born approximation \cite{Born-Wolf.1999} consists in the iterative approximation of a solution of Eq.~(\ref{e:Helmholtz}) of the form
\begin{equation}
  \label{e:Born}
  \hat{\mathbf{E}}(\omega,\mathbf{x}) \sim \hat{\mathbf{E}}_0(\omega,\mathbf{x})
  + \hat{\mathbf{E}}_1(\omega,\mathbf{x}) + \cdots,
\end{equation}
where $\hat{\mathbf{E}}_0$ is a solution of the unperturbed problem (Eq.~(\ref{e:Helmholtz}) with $\delta n_e = 0$) and
\begin{equation}
  \label{e:correctors}
  \nabla \times ( \nabla \times \hat{\mathbf{E}}_j )
  - k_0^2 \hat{\varepsilon}_0 \hat{\mathbf{E}}_j = - k_0^2 \frac{\delta n_e}{n_{e,0}} 
  (\hat{\varepsilon}_0 - I) \hat{\mathbf{E}}_{j-1},
\end{equation}
determines the correctors for $j \geq 1$. Formally at least, the solution for $\hat{\mathbf{E}}_j$ is of order $(\delta n_e / n_{e,0})^j$ so that we may expect convergence of the series for a small-enough fluctuation level, precisely for 
$[\langle \delta n_e^2 /n_{e,0}^2\rangle]^{\frac{1}{2}} \omega_{pe}^2/\omega^2 \ll 1$. When the series converges, the wave energy density averaged over random fluctuations is proportional to
\begin{equation*}
  \langle |\hat{\mathbf{E}} |^2 \rangle \sim |\hat{\mathbf{E}}_0 |^2 +
  \langle \hat{\mathbf{E}}_1^* \cdot  \hat{\mathbf{E}}_1 \rangle + 
  2 \Re \big[
    \hat{\mathbf{E}}_0^* \cdot \langle \hat{\mathbf{E}}_2 \rangle \big] 
  + \cdots,
\end{equation*}
for we have $\langle \hat{\mathbf{E}}_1 \rangle = 0$, which follows from averaging the Eq.~(\ref{e:correctors}) with $j=1$, while in general $\langle \hat{\mathbf{E}}_j\rangle \not = 0$, $j \geq 2$. It is expected therefore that the deviation of $\langle |\hat{\mathbf{E}} |^2 \rangle$ from its unperturbed value $|\hat{\mathbf{E}}_0|^2$ grows quadratically with the fluctuation strength in the Born scattering regime.

The Born expansion~(\ref{e:Born}) has been applied by Karal and Keller~\cite{Karal.1964} in order to obtain an equation for the average wave field $\langle \hat{\mathbf{E}} (\omega, \mathbf{x})\rangle$ and later Mc~Donald~\cite{McDonald.1991} extended their method to derive an equation for the wave-field correlation function $\langle \hat{\mathbf{E}} (\omega, \mathbf{x}) \hat{\mathbf{E}} (\omega, \mathbf{x}')^* \rangle$, from which the radiative transfer model of WKBeam is obtained. Mc~Donald's formal derivation applies to abstract wave equations of the form
\begin{equation}
  \label{e:kk}
  D\psi=0, \qquad D=D_0+D_1 \quad \mbox{with}\quad \langle D_1\rangle=0,
\end{equation}
where $D_0$ is a linear operator acting on a vector $\psi$ in an abstract Hilbert space, and $D_1$ is a linear operator with random coefficients. For the specific problem (\ref{e:Helmholtz}), the wave field $\psi$ is the electric field $\hat{\mathbf{E}}(\omega,\cdot)$, $D_0 \psi$ is the unperturbed operator $\nabla \times (\nabla \times \hat{\mathbf{E}}) - k_0^2 \hat{\varepsilon}_0 \hat{\mathbf{E}}$, and 
$D_1 \psi$ amounts to $- k_0^2 \frac{\delta n_e}{n_{e,0}} (\hat{\varepsilon}_0 - I ) \hat{\mathbf{E}}$ and includes random density fluctuations. The wave field is sought in the form $\psi=\psi_0+\psi_1$, where $\psi_0$ satisfies $D_0\psi_0=0$. After shifting possible singularities in the complex plane~\cite{McDonald.1991}, we can construct an operator $G$ such that $D_0 G = I$. Then Eq.~(\ref{e:kk}) implies 
\begin{equation*}
    \psi = \psi_0 - G D_1\psi,
\end{equation*}
and iterating,
\begin{equation}
  \label{eq:Born2}
  \psi = \psi_0 - GD_1 \psi_0 + G D_1 GD_1 \psi_0 + \cdots,
\end{equation}
which is the Born expansion~(\ref{e:Born}). We can use this series to evaluate the correlation operator $\langle\psi\psi^*\rangle$, and multiplying on the left by $D_0$ we have
\begin{equation*}
  D_0\langle \psi\psi^*\rangle = \langle D_1\langle \psi_0\psi_0^* \rangle D_1\rangle G^*  +\langle D_1GD_1\rangle \langle \psi_0 \psi_0^* \rangle + \cdots.
\end{equation*}
where both $D_0$ and $D_1$ are assumed to be Hermitian and the identity $\langle D_1 \rangle = 0$ has been accounted for. At last, we observe that $\psi_0\psi_0^*$ differs from $\langle \psi\psi^*\rangle$ by second- or higher-order terms, hence, 
\begin{equation}
  \label{e:mcdscatt}
  D_0\langle \psi\psi^*\rangle = \langle D_1\langle \psi\psi^* \rangle D_1\rangle G^*  +\langle D_1GD_1\rangle \langle \psi \psi^* \rangle + \cdots.
\end{equation}
The Weyl symbol of the correlation operator $\langle \psi \psi^*\rangle$ is by definition the average Wigner matrix $W = W(\mathbf{x},\mathbf{N})$ which is a Hermitian-matrix-valued function of position $\mathbf{x}$ and refractive-index vector $\mathbf{N}$. Upon computing the Weyl symbol of the operator equation~(\ref{e:mcdscatt}), the relevant equation for $W$ is readily obtained in a form that depends only on the correlation functions of the random density field, and thus lends itself to the asymptotic solution in the short-wavelength limit in spite of the presence of short scale random fluctuations \cite{McDonald.1988}. 

For the specific case of Eq.~(\ref{e:Helmholtz}), the lowest order approximation of the Wigner matrix $W$ in the short wavelength limit is diagonal on the basis of the two cold plasma polarization vectors and the corresponding two real eigenvalues $w_\alpha$ are referred to as the Wigner functions of the ordinary ($\alpha =$ O) and extra-ordinary ($\alpha =$ X) modes. The dispersion relation imposes the constraint $H_\alpha w_\alpha = 0$ with $H_\alpha = H_\alpha(\mathbf{x}, \mathbf{N})$ being the geometrical optics Hamiltonian for the mode $\alpha$. Then the equation for $W$ reduces to the wave kinetic equation solved by WKBeam, namely,
\begin{equation}
  \label{eq:wke}
  \nabla_N H_\alpha \cdot \nabla_x w_\alpha - \nabla_x H_\alpha \cdot \nabla_N w_\alpha =  S_{\alpha},
\end{equation}
where the Wigner functions $w_\alpha$ are the unknowns and the scattering term $S_\alpha$ can be brought to the form $S_\alpha = \sum_\beta S_{\alpha\beta}$ with
\begin{equation*}
  S_{\alpha\beta} (\mathbf{x},\mathbf{N}) =\int\left[\sigma_{\beta\alpha}(\mathbf{x}, \mathbf{N}',\mathbf{N}) w_\beta(\mathbf{x},\mathbf{N}') - \sigma_{\alpha\beta}(\mathbf{x},\mathbf{N},\mathbf{N}') w_\alpha(\mathbf{x},\mathbf{N})\right]d^3 N',
\end{equation*}
where $\sigma_{\alpha\beta}$ is the scattering cross-section \cite{Weber.2015,Snicker.2018}. Let us remark again that Eq.~(\ref{eq:wke}) holds independently on how short the correlation length and thus spatial scale of the fluctuations themselves is. On the other hand, the Born approximation imposes a limit on the fluctuation amplitude. The Born series~(\ref{eq:Born2}) is controlled by a norm of the operator $D_1$ which for the case of Eq.~(\ref{e:Helmholtz}) can be estimated by 
$[\langle (\delta n_e /n_{e,0})^2 \rangle]^{\frac{1}{2}} \omega_{pe}^2/\omega^2$, hence
the Born approximation should remain valid even for large values of the relative root-mean-square amplitude of the fluctuations if the wave is propagating sufficiently far away from the cut-off\added{, i.e.\ at sufficiently low densities or high frequencies}. Before concluding this section, it is useful to quote a further result derived in \cite{Snicker.2018}, namely that the product $\Sigma_\alpha\Delta\ell$, where $\Delta\ell$ is the distance travelled in the turbulent region, is found to scale as
\begin{equation}
    \Sigma_\alpha\Delta\ell\propto\left({\delta n_e\over n_{e,\subtxt{cut}}}\right)^2 k_0^2L_\perp\Delta\ell,
\end{equation}
where $n_{e,\subtxt{cut}}$ is the cut-off density, $k_0$ is the vacuum wave vector and $L_\perp$ is the perpendicular correlation length of the fluctuations. The quantity $\Sigma_\alpha\Delta\ell$ gives an estimate of the number of scattering events experienced by a given ray.

The code WKBeam has been successfully benchmarked against the paraxial WKB code TORBEAM~\cite{Poli.2001} for different, fusion-relevant scenarios~\cite{Snicker.2018}.

\section{The simulation set-up}\label{s:setup}
The 2D computational domain resembles part of a poloidal cross section in a toroidal magnetic confinement device. The emitting antenna is located in vacuum on the right hand side of the domain. A frequency of $f_0=50\,\mathrm{GHz}$, corresponding to a vacuum wavelength of $\lambda_0\approx6\,\mathrm{mm}$, is chosen for the microwave beam which is described in detail in Sec.~\ref{s:injected_beam}. After a propagation distance of $5\,\mathrm{cm}$ in vacuum, a linearly increasing plasma density profile is encountered, described in detail in Sec.~\ref{s:ne_profile}. The background magnetic field is taken to be homogeneous across the whole domain with a purely toroidal direction and a strength of $B_{\subtxt{tor}} = 1\,\mathrm{T}$, corresponding to a normalized value of $Y = \omega_{ce}/\omega_0 \approx 0.56$. 

\added{Note that the absolute values of the frequency, the plasma density, and of the background magnetic field correspond to the values of the ASDEX-Upgrade tokamak~\cite{Zohm.2007,Manz.2017} reduced by approximately a factor 2.5. The reduced frequency and hence increased vacuum wavelength allows for a coarser numerical grid to be used decreasing the required computational resources. 
Since the electromagnetic wave equation in a cold plasma depends on plasma density and magnetic field through non-dimensional parameters, $X=\omega_{pe}^2/\omega_0^2$ and $Y=\omega_{ce}/\omega_0$ respectively, our simulations models however scenarios at the same $X$ and $Y$ as fusion-relevant scenarios.}

\added{The generation of synthetic density fluctuations is outlined in Sec.~\ref{s:ne_fluctuations}.}

\subsection{The injected microwave beam}\label{s:injected_beam}
The injected beam is Gaussian as being characteristic for typical fusion experiment~\cite{Thumm.2003}. Specifically, for a two-dimensional domain the electric field amplitude of a standard Gaussian beam~\cite{Goldsmith.1998} is given by
\begin{equation}\label{e.Gaussfield}
	E(z,x) = \frac{w_0}{w} \exp \left( -\frac{z^2}{w^2} - ikx - ik\frac{z^2}{2R} + i\phi_0  \right),
\end{equation}
with $z$ the radial distance to the beam axis, $x$ the axial distance to the beam waist, $w_0$ the size of the beam waist, $R$ the radius of curvature of the wavefront, and $\phi_0$ the Gouy phase shift\added{~\cite{Boyd.1980,Feng.2001}}.

Note that $w$, $R$ and $\phi_0$ are all functions of the axial distance $x$ to the beam waist. (Also note that the radial distance is usually denoted with $x$, and the axial distance is usually denoted with $z$.)

In the full-wave code, the field distribution is defined in the antenna plane explicitly by Eq.~(\ref{e.Gaussfield}) and added to the electric field on the grid resembling a \emph{soft source}~\cite{Sullivan.2013}. A focusing beam with the waist located in front of the antenna plane (still inside the computational domain) is considered, and $w$ and $R$ need to be evaluated in the emitting antenna plane using given values of the beam waist $w_0$ and of the axial distance to the waist $x$. In WKBeam, in contrast, the parameters $w$ and $R$ in the antenna plane are direct input parameters. Values of $w = 1.5\,\mathrm{cm}$ and $R = 10\,\mathrm{cm}$ are used in WKBeam, corresponding to $w_0\approx9.7\,\mathrm{mm}$ and $x\approx58.2\,\mathrm{mm}$ for the full-wave code as described in detail in the Appendix.\added{ Those values were chosen to ensure that the simulation domain contains the beam waist and the subsequently diverging beam and still has a reasonable size with respect to the required computational resources.}

\subsection{The electron plasma density profile}\label{s:ne_profile}
\begin{figure}[tb] 
	\centering
	\includegraphics[width=0.45\textwidth]{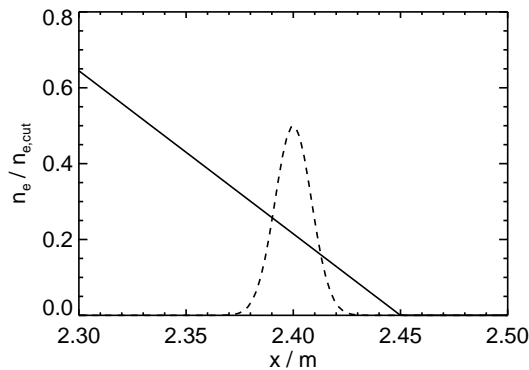}
	\caption{\emph{(Solid line)} Background electron plasma density profile normalized to the O-mode cut-off density and \emph{(dashed line)} the Gaussian envelope for the fluctuation amplitude ($A_0=0.5$, $x_{\subtxt{shift}}=0$, and $w_{\subtxt{fluct}}=0.02\,\mathrm{m}$).}
	\label{f:ne_1D}
\end{figure}

A 1D profile which depends only on the radial coordinate $x$ is chosen for the background electron plasma density. The density starts to increase linearly at $x_{n1}=2.45\,\mathrm{m}$ until $x_{n2}=2.30\,\mathrm{m}$, where a maximum value of $n_{e,\subtxt{max}}=0.2\cdot10^{20}\,\mathrm{m}^{-3}\approx0.65\cdot n_{e,\subtxt{cut}}$ is reached, where $n_{e,\subtxt{cut}}$ refers to the cut-off density of the injected mode which is, if not explicitly stated otherwise, the O-mode. \added{The density values normalized to $n_{e,\subtxt{cut}}$ correspond approximately to those in the scrape-off layer in ASDEX-Upgrade~\cite{Willensdorfer.2012}.} The linear profile is plotted in Fig.~\ref{f:ne_1D} and described by
\begin{equation}\label{e:neProfile}
	n_{e,0}(x)=\left\{ \begin{array}{l}
		n_{e,\subtxt{max}}, \hspace{2.525cm} \mbox{if } x<2.30\,\mathrm{m} \\
		\frac{ n_{e,\subtxt{max}} }{ x_{n1} - x_{n2} } \left( x_{n1} - x \right), \hspace{.5cm} \mbox{if } 2.30\,\mathrm{m}\leq x \leq 2.45\,\mathrm{m} \\
		0, \hspace{3.372cm} \mbox{if } x>2.45\,\mathrm{m}. 
	\end{array}\right.
\end{equation}

\begin{figure}[tb] 
	\centering
	\includegraphics[width=0.45\textwidth]{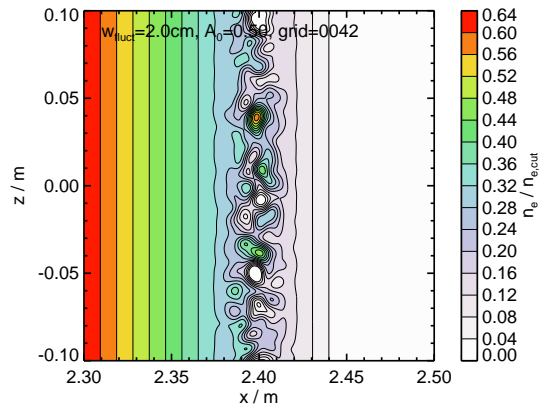}
	\caption{Contour plot of the electron plasma density of one of the samples used as input in the full-wave simulations (for the same parameters as in Fig.~\ref{f:ne_1D}).}
	\label{f:ne_2D}
\end{figure}

A layer of turbulent density fluctuations is then added to the background profile, as described in Sec.~\ref{s:ne_fluctuations}. The Gaussian envelope of the fluctuation amplitude is given by the expression
\begin{equation}\label{e:fluct_envelope}
	F(x) = A_0\cdot\exp\left\{ - \frac{ \left[ ((x+x_{\subtxt{shift}})-R_0)/a - 1.25 \right]^2 }{w_{\subtxt{fluct}}^2} \right\},
\end{equation}
where $A_0$ is the normalized fluctuation strength, $R_0=1.65\,\mathrm{m}$ and $a=0.6\,\mathrm{m}$ correspond respectively to the major and minor radius, and $w_{\subtxt{fluct}}$ defines the width of the Gaussian envelope. The parameter $x_{\subtxt{shift}}$ is used to shift the fluctuation layer radially, where a value of $x_{\subtxt{shift}}=0$ corresponds to the center of the layer being located at $x=2.40\,\mathrm{m}$. Note that the values used for major and minor radii correspond to the ASDEX Upgrade tokamak~\cite{Manz.2017}. Figure~\ref{f:ne_2D} shows as an example one sample for the actual 2D density profile used in the full-wave simulations. The finite spatial extent of the fluctuations in $x$-direction can be clearly seen. It is remarked that in the region around $x_{\subtxt{shift}}=0$ the ratio between plasma density and cut-off density closely matches the corresponding value expected in the ITER standard scenario~\cite{Snicker.2018}.

To ensure statistical significance of the full-wave simulations, the ensemble of density profiles needs to be large enough for each set of turbulence parameters. A size of $N=3000$ turbulence realizations has been found to yield relevant results as will be demonstrated in Sec.~\ref{s:A0_scan_Omode}. In WKBeam on the other hand, the statistical parameters of the turbulent density fluctuations are used as an input to directly calculate the \emph{average effect} on the microwave beam. This leads to a significant reduction of computational time as compared to full-wave simulations. 

The perpendicular correlation length of the density structures is set to a value of $L_\perp \approx 5\,\mathrm{mm}$ which is close to the vacuum wavelength of the injected microwave ($\lambda_0 \approx 6\,\mathrm{mm}$). According to Ref.~\cite{Koehn.2016}, this can result in pronounced scattering of the microwave. The correlation length is predicted to scale like $L_\perp \approx 5 - 10 \rho_s$~\cite{Rhodes.2002}, with the drift scale parameter $\rho_s = \sqrt{T_e m_i}/(eB_0)$~\cite{Ramisch.2005} (where $T_e$ is the electron temperature and $m_i$ the ion mass). Assuming typical values for an ASDEX Upgrade discharge~\cite{Fischer.2010, Nold.2010} with deuterium ions, the chosen value of $L_\perp=5\,\mathrm{mm}$ lies within that range.

\subsection{Density fluctuations\added[remark={Note that this subsection has been moved entirely from the section ''Numerics'' as suggested by one of the referees}]{}}\label{s:ne_fluctuations}
In order to study the effect of plasma density fluctuations on a traversing microwave beam by means of full-wave simulations it is necessary to let the microwave beam interact with an \emph{ensemble} of density profiles, each being a sample of the same random field, and then average over the resulting wave electric fields. To generate a large number of individual samples, we use synthetic turbulence, as it allows us to generate the required large ensembles in a reasonable time as opposed to using large-scale plasma turbulence codes. It also allows us to ensure that the statistics of the random field is the same as that assumed in the WKBeam code.

The computational domain is defined on a 2D grid which is a reasonable simplification as turbulence in magnetized plasmas is highly anisotropic with typically very small wave numbers parallel to the background magnetic field~\cite{Zweben.2007}. The 2D domain corresponds approximately to a poloidal cross section in a toroidal magnetic confinement device. The full electron plasma density profile in the 2D simulation domain, described in detail in Sec.~\ref{s:ne_profile}, can be written as
\begin{equation}\label{e:ne_profile_2D}
	n_e (x,z) = n_{e,0}(x) \left( 1 + F(x)\,\delta n(x,z) \right),
\end{equation}
where $x$ and $z$ are the radial and vertical coordinates, respectively, $n_{e,0}(x)$ is the unperturbed background profile, $F(x)$ an envelope of the fluctuations' amplitude basically defining their spatial location, and $\delta n(x,z)$ is a random field such that $\delta n_e (x,z)/ n_{e,0}(x) = F(x) \delta n(x,z)$ is the density fluctuation. Note that there is no dependence on time here as the density fluctuations appear to be frozen in the time frame of the microwave (also referred to as \emph{frozen plasma assumption}): typical frequencies of the density fluctuations lie in the kHz range~\cite{Zweben.2007}, whereas the microwave oscillates in the GHz range. In addition, for the densities considered here, the group velocity of the microwave is several orders of magnitude above the propagation speed of the density structures~\cite{Zweben.2007}. 

The fluctuations themselves are generated by a truncated sum of Fourier-like modes:
\begin{equation}\label{e:ne_profile_fluctuations}
	\delta n(x,z) = \sum_i^{M_i} \sum_j^{M_j} A_{i,j} \cos\left[ k_{x,i}x + k_{z,j}z + \varphi_{i,j} \right],
\end{equation}
with $A_{i,j}$ the amplitudes of the modes and $\varphi_{i,j}$ independent random phases uniformly distributed in the interval $\left[0,2\pi\right)$. The correlation length of the two-point auto-correlation function of the density fluctuations correspond to the average perpendicular structure size $L_\perp$. Although the corresponding spectra in the experiments exhibit usually some kind of power law (see e.g.\ Refs.~\cite{Stroth.2004,Conway.2008}) in contrast to the Gaussian shape used here, this is not expected to lead to significantly different scattering of the microwave beam for the parameters used here as the power laws differ most strongly from the Gaussian at large $k$-values for which, according to previous investigations~\cite{Koehn.2016}, strongly reduced scattering is expected.

In WKBeam, the effect of plasma density fluctuations is included via a scattering operator. The input parameters in the current model can be reduced to the spatial localization of the fluctuation layer, $F(x)$, and the two correlation lengths, $L_\perp$ and $L_{||}$ (for details, see Ref.~\cite{Snicker.2018}). This ensures that both codes use the same plasma density profiles (including fluctuations) as input.

\section{Data analysis}\label{s:analysis}
The full-wave simulations, based on a time-dependent scheme, start with the excitation of the microwave beam. They are stopped when a steady state solution is achieved. Including a safety margin in computational time, this corresponds to a value of $T=100\,T_{\subtxt{wave}}$, where $T_{\subtxt{wave}}$ denotes the oscillation period. At various radial positions, the time-averaged squared wave electric field is recorded across the whole $z$-range of the computational domain: 
\begin{equation}
	\tilde{E}^2=\frac{1}{T}\,\sum_t \tilde{E}_x^2 + \tilde{E}_y^2 + \tilde{E}_z^2,
\end{equation}
where $t$ is the time coordinate and the tilde indicates that a scenario with turbulent density fluctuations is used (as opposed to a scenario without fluctuations where just the linear profile as described by Eq.~(\ref{e:neProfile}) is used).
Such detector antenna signals are acquired for each sample at a given set of radial positions $x$. The \emph{ensemble-averaged} signals of the full-wave simulations can then be compared with the output of WKBeam (which yields directly the squared wave electric field). 

As will be demonstrated in the following section, the ensemble-averaged beam cross-section can be approximately described by a Gaussian. In order to quantify the broadening of the injected microwave beam, it is thus convenient to fit a Gaussian of the shape
\begin{equation}\label{e:fit_gauss}
	f(z)=a_0\exp\left\{ -\left(\frac{z-a_1}{2a_2}\right)^2 \right\}
\end{equation}
to the (ensemble-averaged) scattered signal in the detector antenna planes with $a_0$, $a_1$, and $a_2$ being the fit parameters. The value obtained for the (averaged) beam width $\bar{w}$, corresponding to the fit parameter $a_2$, is then compared with the width $w$ of the same beam propagating in the unperturbed scenario (i.e. without fluctuations but with the background profile). A \emph{normalized beam broadening} $b=\bar{w}/w$ is obtained in this way for each set of turbulence parameters allowing to compare and benchmark the WKBeam results with the full-wave simulations. Note that another example for FDTD full-wave simulations of electromagnetic waves passing through random media consists in calculating the scattering coefficient, see e.g.\ Ref.~\cite{Capoglu.2009}. The relevant physical quantity in our case is however the beam broadening as outlined in the introduction. 

Although the transverse ensemble-averaged beam profile can be well approximated by a Gaussian in most cases, there will be a few scenarios where a Cauchy distribution is more suitable (as will be discussed in the following Sections). Therefore, a general Cauchy distribution of the form
\begin{equation}\label{e:fit_cauchy}
	f(z) = a_0 \frac{1}{\pi} \frac{a_1}{ (z-a_2)^2 + a_1^2 }
\end{equation}
will also be fitted to the detector antenna signals (via a non-linear least square fit), where $2a_1$ corresponds to the full width at half maximum (FWHM) and $a_2$ to the median ($a_0$ fits the amplitude). The normalised beam broadening for those cases will be analysed in terms of the FWHM, i.e.\ the value of $a_1$ obtained from the ensemble-averaged signal is normalised with the corresponding value for the case without fluctuations.

\added{The plasma density in the fluctuation layer can locally reach values close to the cut-off density of the microwave if large fluctuation levels are considered. This can result in microwave power being reflected. The corresponding quantity is routinely measured in the full-wave simulations and its value, averaged over the whole ensemble of fluctuations, is found to be at maximum $1\,\%$ for the worst case scenario, and at least one order of magnitude lower for most cases.}

\section{Influence of the fluctuation level on beam broadening}\label{s:A0_scan_Omode}
In this section, the results from full-wave simulations and WKBeam calculations are compared first for the case without turbulent density fluctuations\added{ (the zeroth case, comparison in vacuum, yielded excellent agreement and is not included in this paper)}. For this case we find that the two codes are indeed in very good agreement. Having established a reference solution, density fluctuations are included with their envelope localized at a radial position, as described in Sec.~\ref{s:ne_profile}. The fluctuation amplitude is varied in a series of scans and the resulting values for the beam broadening are compared for the two codes.

\subsection{Case without turbulence}
\begin{figure}[tb] 
	\centering
	\includegraphics[width=0.45\textwidth]{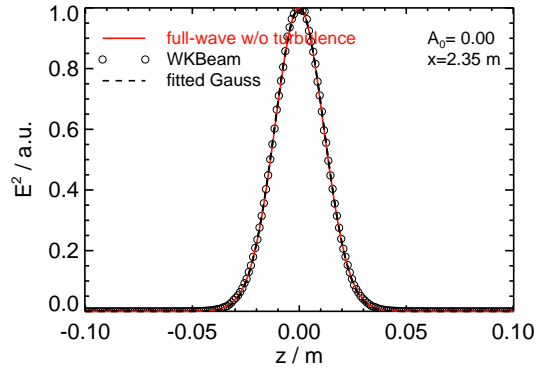}
	\caption{Detector antenna signals at a radial position of $x=2.35\,\mathrm{m}$ from full-wave simulations and WKBeam calculations with Gaussians fitted to them for the case without turbulence (but with the background density profile as described by Eq.~(\ref{e:neProfile})).}
	\label{f:detAnt15_noTurb_Omode}
\end{figure}

The linear density profile given by Eq.~(\ref{e:neProfile}) is taken in this scenario without any fluctuations, i.e. $A_0=0$. The resulting detector antenna signals at a radial position of $x=2.35\,\mathrm{m}$ from both, the full-wave simulations and the WKBeam calculations, agree within the resolution of the plot shown in Fig.~\ref{f:detAnt15_noTurb_Omode}. 
The Gaussians fitted to the signals (see Eq.~(\ref{e:fit_gauss})) are also included. The value of the beam size $w$ obtained from the Gaussian for WKBeam is larger by $0.2\,\%$ than the corresponding value for the full-wave simulations. It is in principle possible to further reduce this difference by adjusting the spatial resolution in both codes since the detector antenna signals need to be interpolated to the respective other code. This would however increase the total demand for computational resources and the introduction of the turbulent fluctuations leads anyway to an increase of at least an order of magnitude in the difference (see Sec.~\ref{s:first_turbulence_example}). We thus decided to use the values yielding the (already very good) agreement shown in Fig.~\ref{f:detAnt15_noTurb_Omode}.

\subsection{Including a layer of turbulent density fluctuations}\label{s:first_turbulence_example}
\begin{figure}[tb] 
	\centering
	\includegraphics[width=0.45\textwidth]{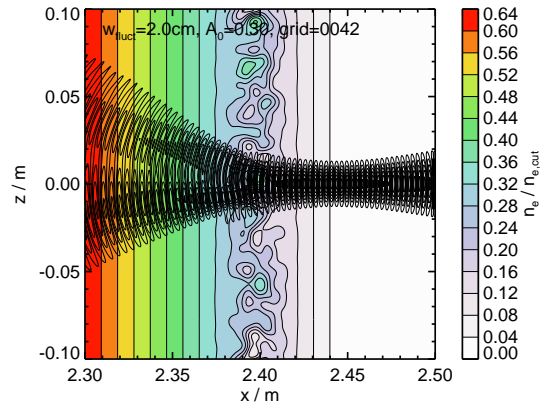}
	\caption{Electron plasma density and snapshot of the absolute value of the wave electric field as obtained from full-wave simulations for one sample ($A_0=0.30$, $w_{\subtxt{fluct}}=2\,\mathrm{cm}$, $x_{\subtxt{shift}} = 0$, see Eq.~(\ref{e:fluct_envelope})).}
	\label{f:fullwave__example__scenario08.16}
\end{figure}
As a next step, a layer of turbulent plasma density fluctuations is added to the background profile. The position and the width of the layer are kept constant, using values of $x_{\subtxt{shift}} = 0$ and $w_{\subtxt{fluct}}=2\,\mathrm{cm}$, respectively. To illustrate the effect of the fluctuations, a snapshot of the absolute wave electric field obtained from full-wave simulations for a single sample is shown in Fig.~\ref{f:fullwave__example__scenario08.16}. The fluctuation layer clearly perturbs the injected beam, leading to a splitting into multiple beams which destroys the intended spatial localization of the absorption.

\begin{figure}[tb] 
	\centering
	\includegraphics[width=0.45\textwidth]{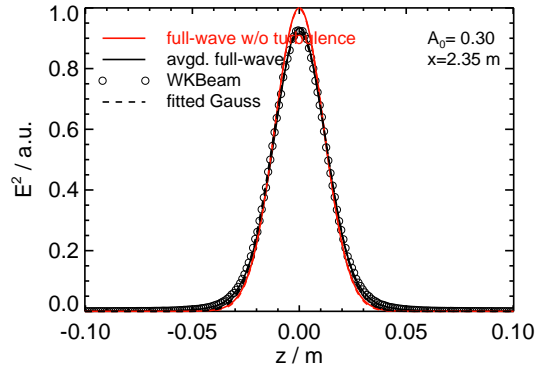}
	\caption{Same as Fig.~\ref{f:detAnt15_noTurb_Omode}, but for the case with turbulent density fluctuations ($A_0=0.30$, $w_{\subtxt{fluct}}=2\,\mathrm{cm}$, $x_{\subtxt{shift}} = 0$, see Eq.~(\ref{e:fluct_envelope})).}
	\label{f:detAnt15_Omode_shift0_wfluct2_A030}
\end{figure}

As an example, full-wave and WKBeam beam profiles are compared for a fluctuation amplitude $A_0=0.3$ in Fig.~\ref{f:detAnt15_Omode_shift0_wfluct2_A030}. On average, i.e.\ averaging over the full ensemble, a small broadening of the beam as compared to the case without turbulence is found. The ensemble-averaged signal resembles a smooth Gaussian-like beam, illustrating the sufficient size of the ensemble. Note that the signal is symmetric around $z=0$ and thus does on average not change its original direction of propagation. Comparing with WKBeam, no differences are noticeable in this representation, proving the validity of the WKBeam calculations for this set of parameters. A tiny deviation to a Gaussian fit, also included in the figure, is only found in the tail of signals, which is slightly elevated. The elevated tails can become more pronounced, as will be discussed in Sec.~\ref{s:shift_scan}. From the full-wave simulations, a broadening \added{normalized to the case without fluctuations} of $b_{\subtxt{fw}}=1.070\added{\,(\pm0.002)}$ is obtained and for the WKBeam calculations it is $b_{\subtxt{WB}}=1.086$ which is larger by approximately $1\,\%$. \added{Normalizing the broadening to the case without plasma (i.e.\ the vacuum case), $b_{\subtxt{fw}}$ and $b_{\subtxt{WB}}$ need to be multiplied with an additional factor of approximately $1.119$, resulting in $b_{\subtxt{fw,vac}}\approx1.197$ and $b_{\subtxt{WB,vac}}\approx1.215$. If not mentioned explicitly otherwise, the beam broadening is normalized to the case without fluctuations (but with plasma) in the rest of this paper.}

\subsection{Scanning the fluctuation level}
\begin{figure}[tb] 
	\centering
	\includegraphics[width=0.45\textwidth]{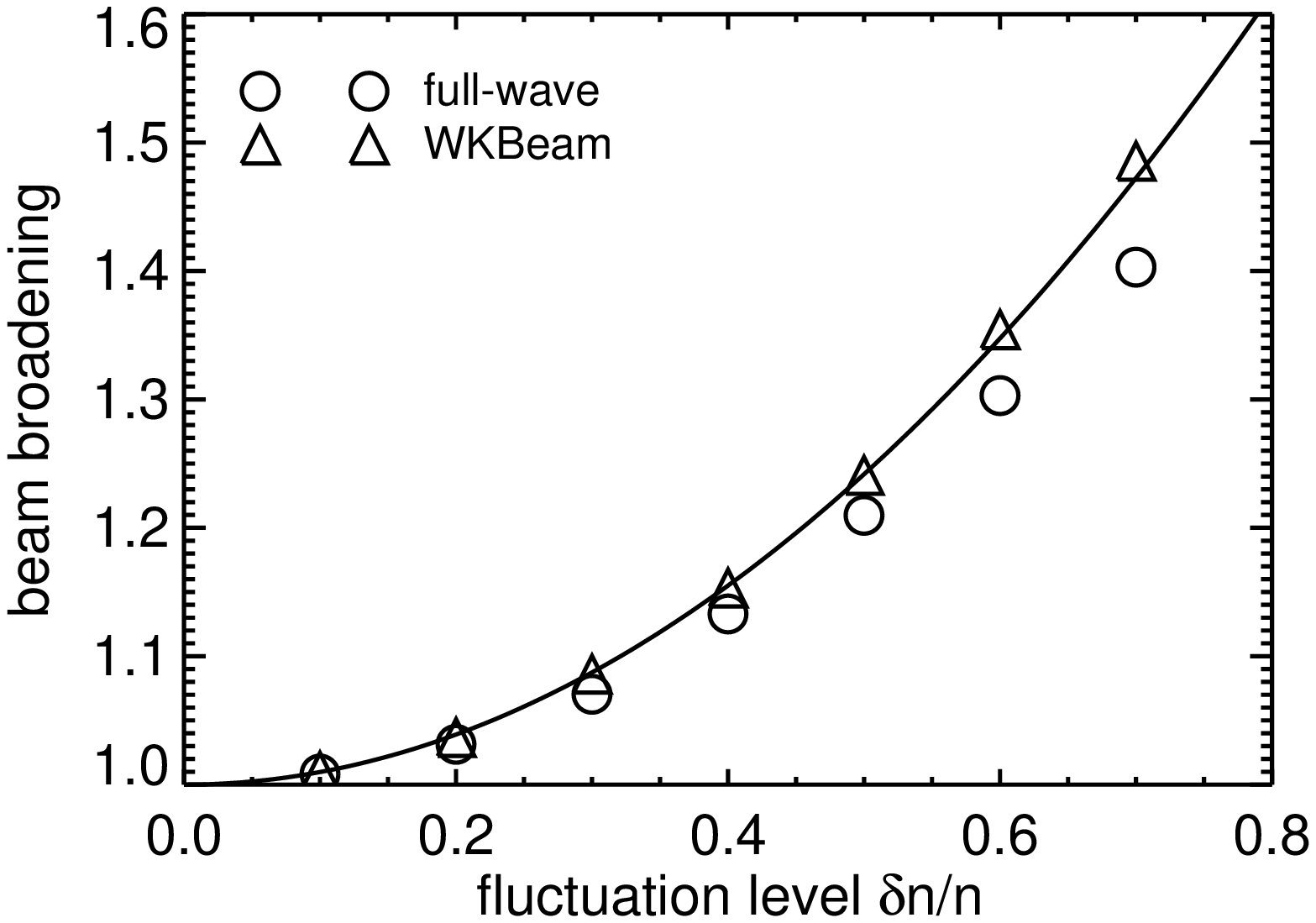}
	\includegraphics[width=0.45\textwidth]{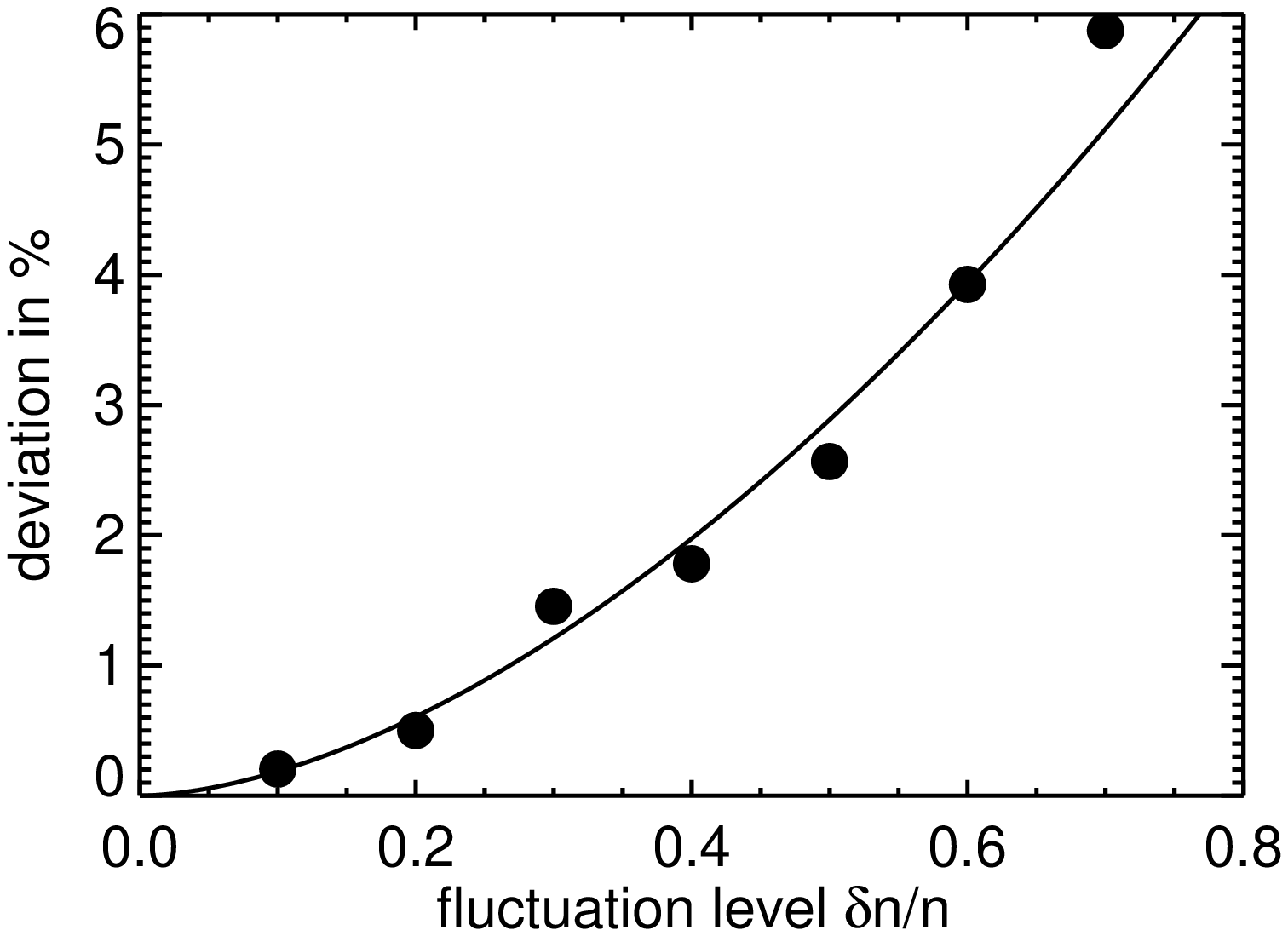}
	\caption{\emph{(Left)} Beam broadening as a function of the fluctuation amplitude at a detector antenna position of $x=2.35\,\mathrm{m}$ from full-wave simulations and WKBeam calculations as indicated in the plot.  \emph{(Right)} Percentage deviation of WKBeam to full-wave results using the values shown in the left plot.}
	\label{f:beambroadening_Omode_xshift0_wfluct2}
\end{figure}

One of the main goals of this paper is to investigate the deviation of the beam broadening as predicted by the WKBeam code with respect to the reference solution provided by the full-wave solver. In particular, as discussed in Sec.~\ref{s:WKBeam}, the Born approximation is expected to become inaccurate with increasing fluctuation level and background density. Figure~\ref{f:beambroadening_Omode_xshift0_wfluct2} \emph{(left)} shows the resulting scaling which yields an increased broadening with increasing fluctuation level. \added{Error bars are not shown since the standard deviation of the ensemble-averaged beam broadening is smaller than the symbol size used in the plots.}

The broadening for WKBeam is consistently larger than the full-wave solution and the absolute difference increases with increasing fluctuation strength. In WKBeam, the beam broadening $b$ as a function of the fluctuation amplitude $A_0$ follows a power law. The solid line included in the plot in Fig.~\ref{f:beambroadening_Omode_xshift0_wfluct2} \emph{(left)} corresponds to a fit to the WKBeam values, obtaining a functional dependence of $b = 0.96\cdot A_0^{1.99}$. For small fluctuation levels, a quadratic dependence is expected as the scenario resembles a \emph{phase screen}~\cite{Andrews.2005}. The full-wave simulations, in contrast, exhibit a reduced increase for large values of $A_0$. WKBeam is thus overestimating the broadening for large fluctuation levels which can be illustrated by the relative difference plotted in Fig.~\ref{f:beambroadening_Omode_xshift0_wfluct2} \emph{(right)}. The maximum overestimation is with $6\,\%$ still considered to be small. It can become more significant if the background density is larger, as presented in the following section.

\section{Influence of the \replaced{turbulence radial location}{background density} on beam broadening}\label{s:shift_scan}
In this section, the background density in the fluctuation layer is varied by shifting the layer radially, i.e.\ along the $x$-direction. To this end, the parameter $x_{\subtxt{shift}}$ (see Eq.~(\ref{e:fluct_envelope})) is varied in the range $x_{\subtxt{shift}} = 0,\ldots,4\,\mathrm{cm}$ in steps of $1\,\mathrm{cm}$, corresponding to background density values at the center of the fluctuation layer of $n_e/n_{e,\subtxt{cut}}=0.21,0.25,0.30,0.34,0.38$, respectively. The shape of the fluctuation envelope is thus not varied, only its radial position. 

\begin{figure}[tb] 
	\centering
	\includegraphics[width=0.45\textwidth]{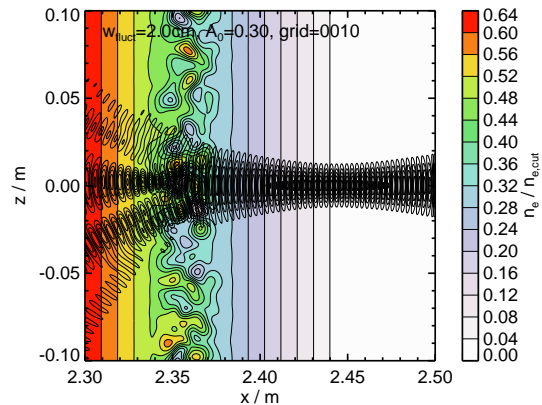}
	\caption{Same as Fig.~\ref{f:fullwave__example__scenario08.16}, but with the turbulence layer shifted to the left ($A_0=0.30$, $w_{\subtxt{fluct}}=2\,\mathrm{cm}$, $x_{\subtxt{shift}} = 4\,\mathrm{cm}$, see Eq.~(\ref{e:fluct_envelope})).}
	\label{f:fullwave__example__scenario08.52}
\end{figure}

Figure~\ref{f:fullwave__example__scenario08.52} shows as an example a full-wave simulation for a value of $x_{\subtxt{shift}}=4\,\mathrm{cm}$ and $A_0=0.30$ with the fluctuation layer centered around a position of $x=2.36\,\mathrm{m}$. The increased background density at the fluctuation layer is expected to result in stronger scattering as refraction effects become more pronounced: power scattered by the turbulent density structures off the original direction of propagation experiences stronger refraction resulting on average in an increased beam broadening.

\begin{figure}[tb] 
	\centering
	\includegraphics[width=0.45\textwidth]{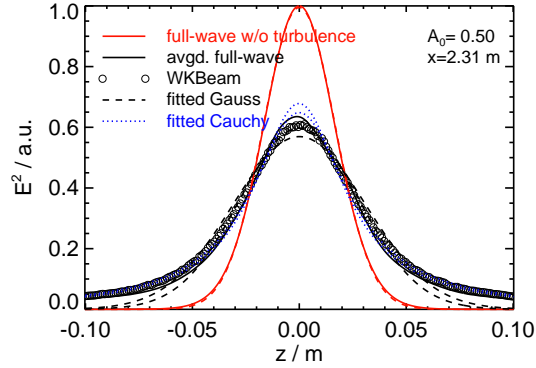}
	\caption{Detector antenna signals at a radial position of $x=2.31\,\mathrm{m}$ from full-wave simulations and WKBeam calculations with Gaussians and Cauchy distributions fitted to them for $A_0=0.50$ and $x_{\subtxt{shift}}=4\,\mathrm{cm}$.}
	\label{f:detAnt19_Omode_shift4_wfluct2_A050}
\end{figure}

As a further example, the detector antenna signals for $x_{\subtxt{shift}}=4\,\mathrm{cm}$ and $A_0=0.50$ are plotted in Fig.~\ref{f:detAnt19_Omode_shift4_wfluct2_A050}. Since the antenna position used in the previous section ($x=2.35\,\mathrm{m}$) would be situated inside the fluctuation layer, a position of $x=2.31\,\mathrm{m}$ is chosen here. As a first observation, the ensemble-averaged full-wave signal and the WKBeam signal are both broader (and correspondingly with a reduced peak amplitude) than the example shown in Fig.~\ref{f:detAnt15_Omode_shift0_wfluct2_A030}, where \emph{(a)} the background density was lower and \emph{(b)} the normalized fluctuation amplitude was with a value of $A_0=0.3$ also lower. Although both these differences make the scenario considered here a harder test for the Born approximation, the agreement between WKBeam and full-wave solution is still remarkably good.

Although a disagreement between full-wave and WKBeam signals can be seen, it is not considered to be significant. 

Another observation is related to the shape of the signals: a Gaussian seems no longer be the adequate function to describe them, a pronounced elevation at the tails can be clearly seen. Therefore, a Cauchy distribution as described by Eq.~(\ref{e:fit_cauchy}) has also been fitted to the signals. Comparing the fitted Cauchy distribution with the fitted Gaussian, see Fig.~\ref{f:detAnt19_Omode_shift4_wfluct2_A050}, the first one seems to be more suitable to describe the broadened microwave beam for these fluctuation parameters. This finding corresponds to the results presented in Ref.~\cite{Snicker.2018}: a Cauchy distribution thus corresponds to a scattering process of super-diffusive nature, whereas a Gaussian shape corresponds to a diffusive process. 

Note that the beam broadening deduced from either the Gaussian or the Cauchy does not differ much: for the Gaussian fit it is \replaced{$b_{\subtxt{fw}}=1.652\,(\pm0.001)$}{$b_{\subtxt{fw}}\approx1.65$} and \replaced{{$b_{\subtxt{WB}}=1.838$}}{$b_{\subtxt{WB}}\approx1.84$} for full-wave and WKBeam, respectively, and for the Cauchy fit the values are $b_{\subtxt{fw}}\approx1.60$ and $b_{\subtxt{WB}}\approx1.74$. \added{To get the broadening normalized to the vacuum case a slightly different factor as in Sec.~\ref{s:A0_scan_Omode} is required here due to the different detector antenna position: the factor is approximately $1.211$, resulting in $b_{\subtxt{fw,vac}}\approx2.00$ and $b_{\subtxt{WB,vac}}\approx2.23$ for the Gaussian fit.}

The broadening is no longer the only measure of interest. Instead, an increasing amount of energy is located in the tails. This could create a problem for an actual experiment, as it basically means that more scattering events far off the original direction of beam propagation occur, threatening diagnostics or other wall components.

\begin{figure}[tb] 
	\centering
	\includegraphics[width=0.45\textwidth]{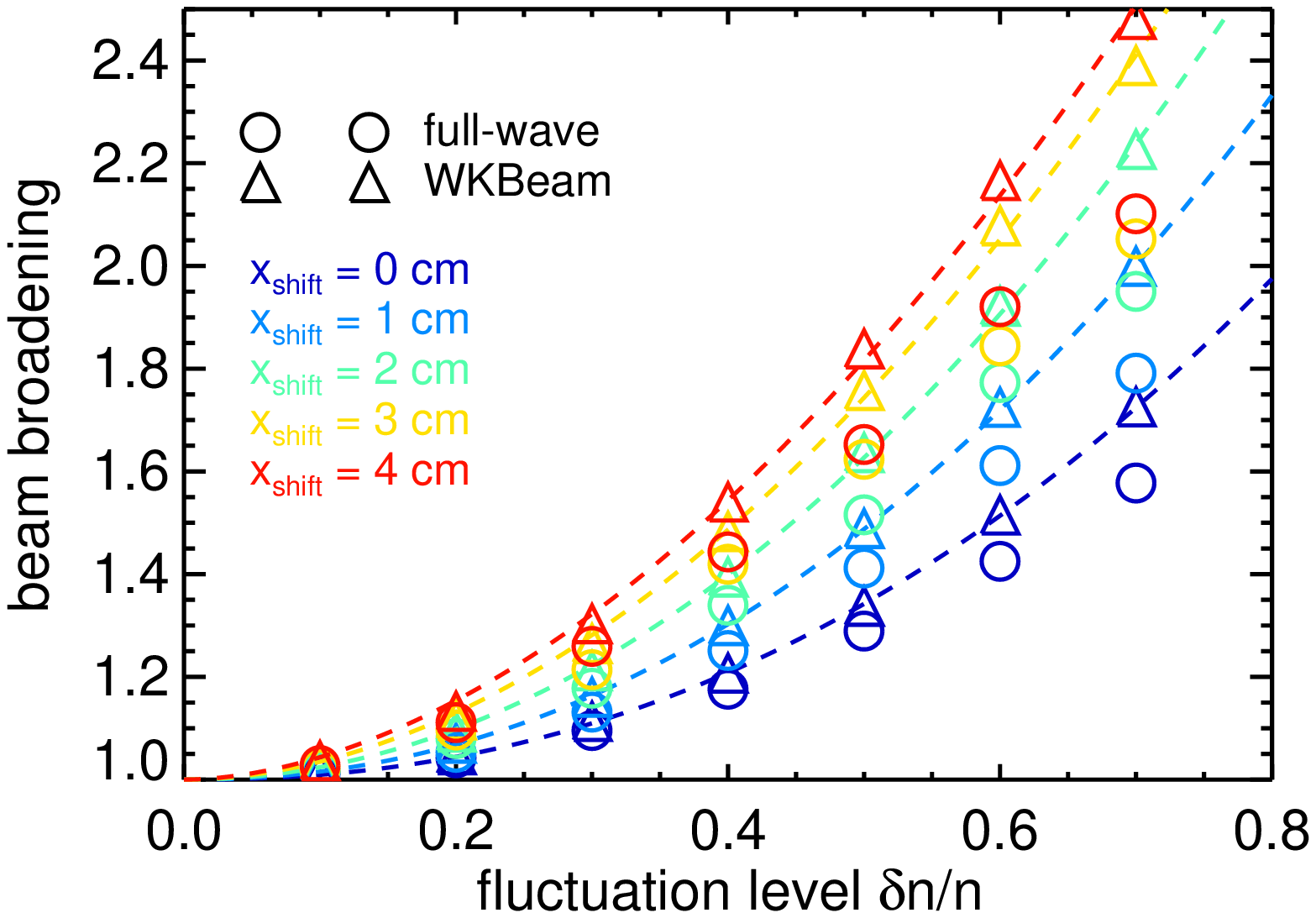}
	\includegraphics[width=0.45\textwidth]{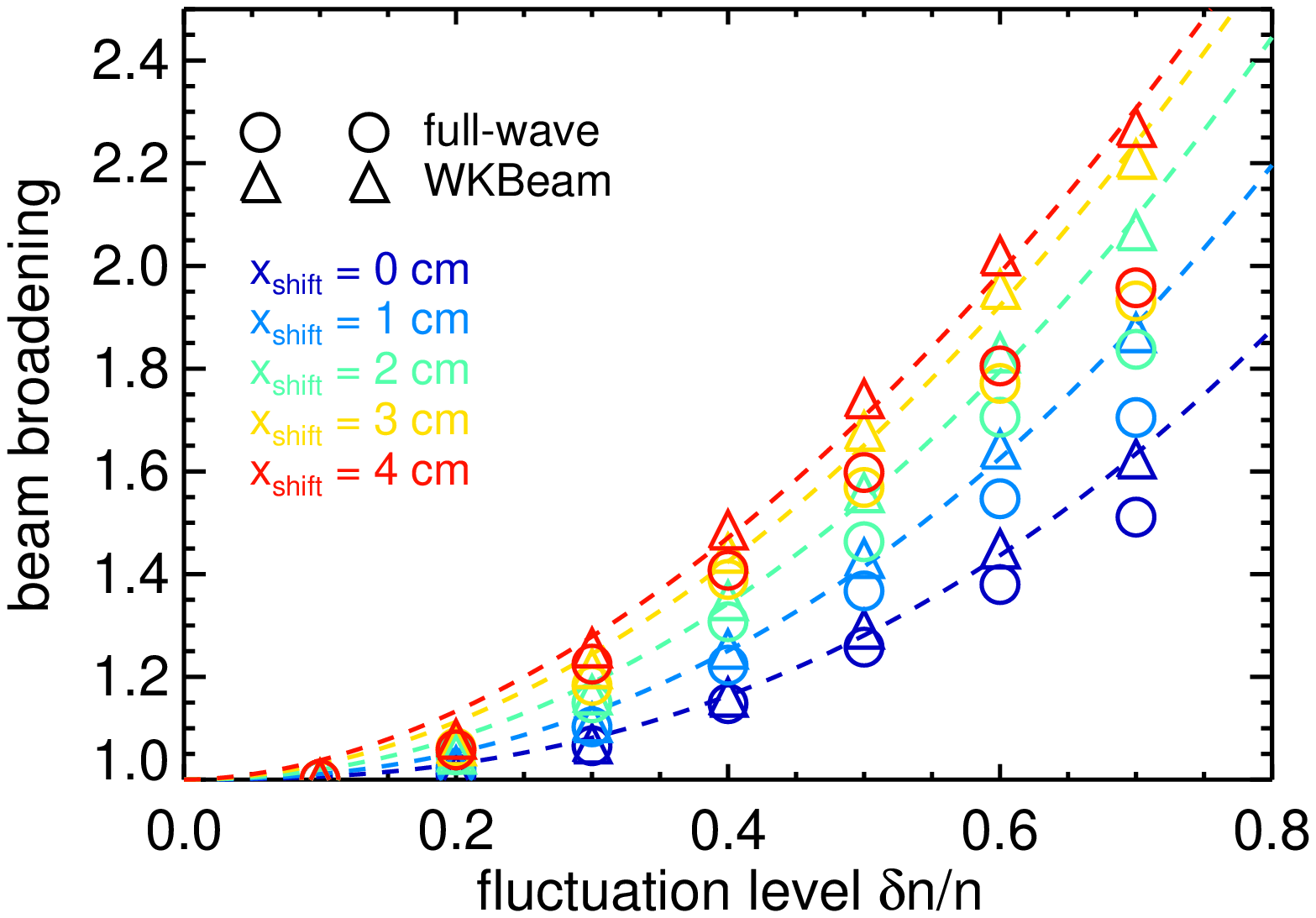}
	\caption{Beam broadening as a function of the fluctuation amplitude at a detector antenna position of $x=2.31\,\mathrm{m}$ from full-wave simulations and WKBeam calculations as deduced from \emph{(left)} Gaussians and \emph{(right)} from Cauchy distributions fitted to the detector antenna signals. Power laws are fitted to the WKBeam results (dashed lines).}
	\label{f:beambroadening_Omode_xshiftScan_gaussw_detAnt19}
\end{figure}

The resulting beam broadening deduced from the Gaussians and the Cauchy distributions fitted to the detector antenna signals is shown, respectively, in Fig.~\ref{f:beambroadening_Omode_xshiftScan_gaussw_detAnt19} \emph{(left)} and \emph{(right)} as a function of the fluctuation amplitude $A_0$ with $x_{\subtxt{shift}}$ as additional parameter. The plot shows the clear trend of increased beam broadening with increasing values of $x_{\subtxt{shift}}$ (and increasing values of $A_0$) for both codes. Significant broadening by more than a factor of two is found. One can also see that the WKBeam values can still be represented by power laws and that the full-wave values are consistently smaller. Not much differences can be seen between using a Gaussian or the Cauchy distribution in this representation, the broadening seems to be very similar. An asymptotic behavior is observed towards larger values of $x_{\subtxt{shift}}$, i.e.\ higher background densities: the slope of the fitted power law for $x_{\subtxt{shift}}=4\,\mathrm{cm}$ is only slightly larger than for $x_{\subtxt{shift}}=3\,\mathrm{cm}$ whereas there is a substantial difference going from $x_{\subtxt{shift}}=0\,\mathrm{cm}$ to $x_{\subtxt{shift}}=1\,\mathrm{cm}$.

\begin{figure}[t] 
	\centering
	\includegraphics[width=0.45\textwidth]{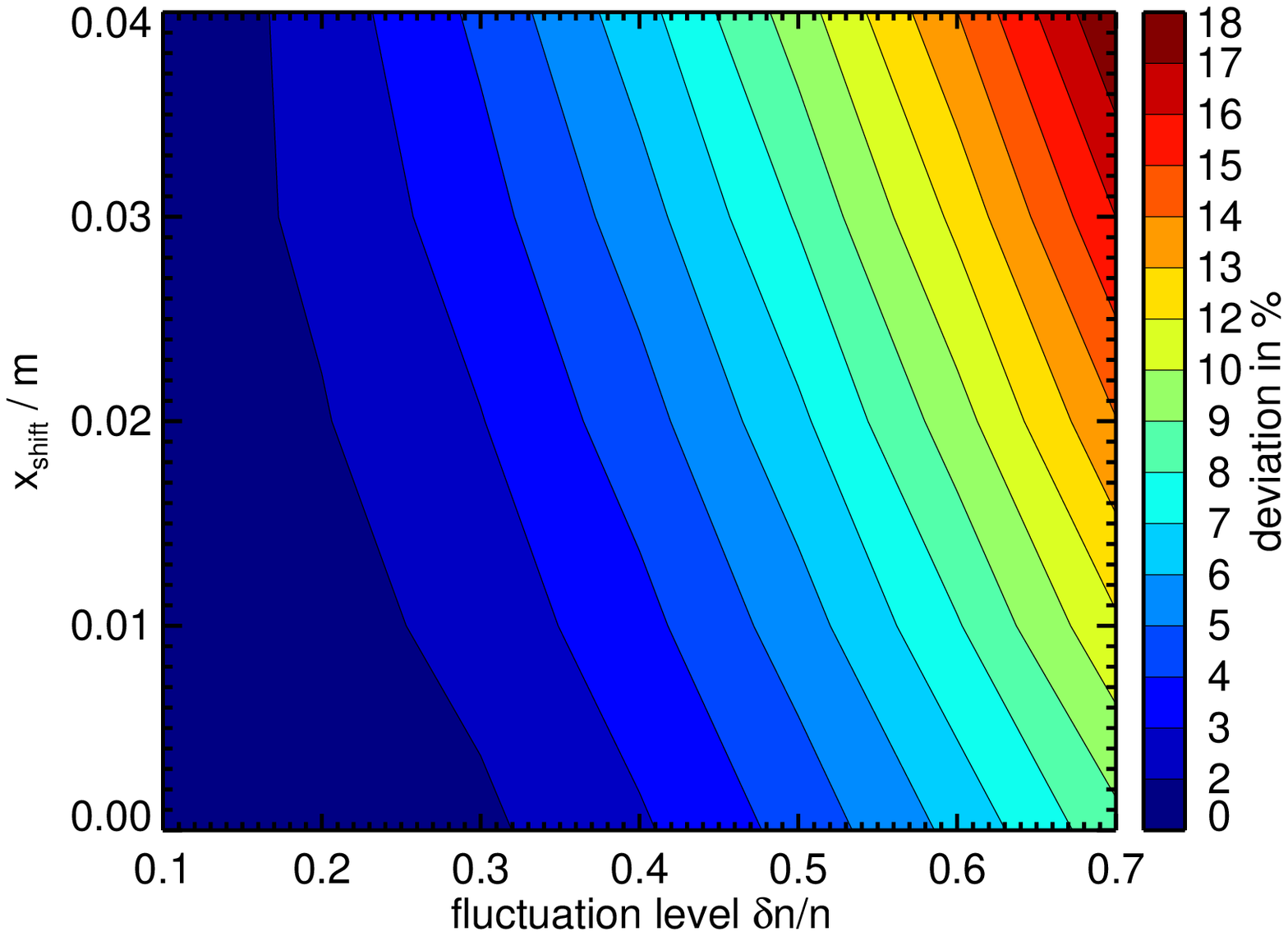}
	\includegraphics[width=0.45\textwidth]{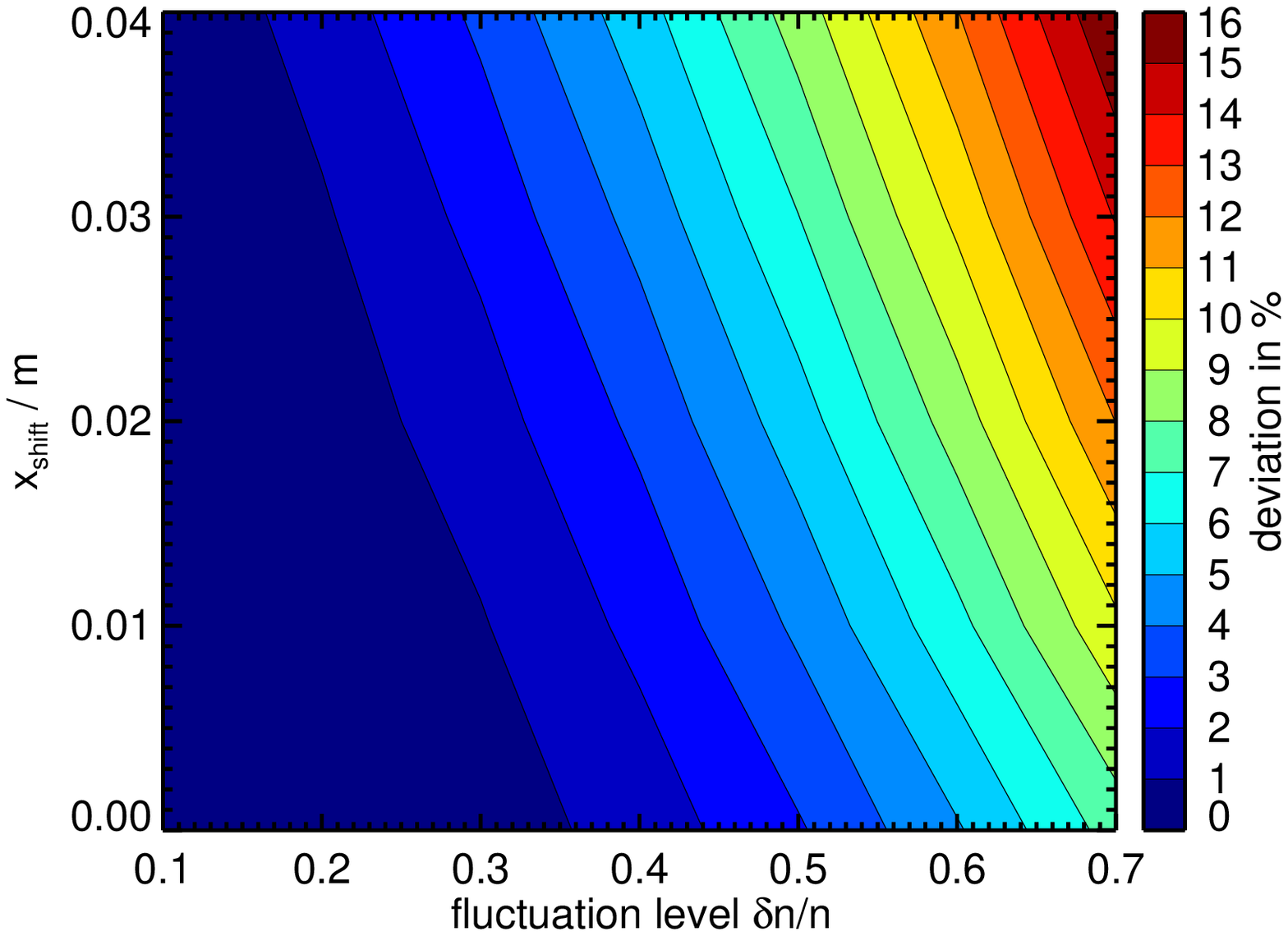}
	\caption{Polynomial fit to the deviation of beam broadening obtained from fitting \emph{(left)} a Gaussian ($\chi^2\approx2.3$ using all 35 beam broadening values) and \emph{(right)} a Cauchy distribution ($\chi^2\approx1.6$ using all 35 beam broadening values) to detector antenna signals at $x=2.31\,\mathrm{m}$ for WKBeam calculations with respect to full-wave simulations (i.e.\ overestimation of WKBeam calculations compared to the full-wave simulations).}
	\label{f:beambroadening_Omode_xshiftScan_cauchyb_deviation_contPlot_fit2D}
\end{figure}

The overestimation of the beam broadening of WKBeam can be described in a more quantitative way by calculating the percentage deviation $d$ of the WKBeam values to the full-wave values. Polynomials of 2\textsuperscript{nd} order can be fitted to the deviation as a function of fluctuation level $A_0$ and $x_{\subtxt{shift}}$ (representing the background density). As shown in Fig.~\ref{f:beambroadening_Omode_xshiftScan_cauchyb_deviation_contPlot_fit2D}, maximum deviations of $18\,\%$ are found for the parameters used in this paper, where for fluctuation levels below $50\,\%$ the overestimation of WKBeam stays below $10\,\%$.

%
%
%

\section{Influence of the width of the fluctuation layer on beam broadening}\label{s:wfluctScan}
The width of the fluctuation layer is varied in this section in order to investigate the influence of the propagation length inside of the fluctuating density area on beam broadening. The parameter $w_{\subtxt{fluct}}$, see Eq.~(\ref{e:fluct_envelope}), is varied, where values of $w_{\subtxt{fluct}}=1,2,3\,\mathrm{cm}$ are used. The parameter $x_{\subtxt{shift}}$ is kept constant at a value of $x_{\subtxt{shift}}=0$ and the fluctuation amplitude $A_0$ is varied over the same range as in the previous cases. 
\begin{figure}[tb] 
	\centering
	\includegraphics[width=0.45\textwidth]{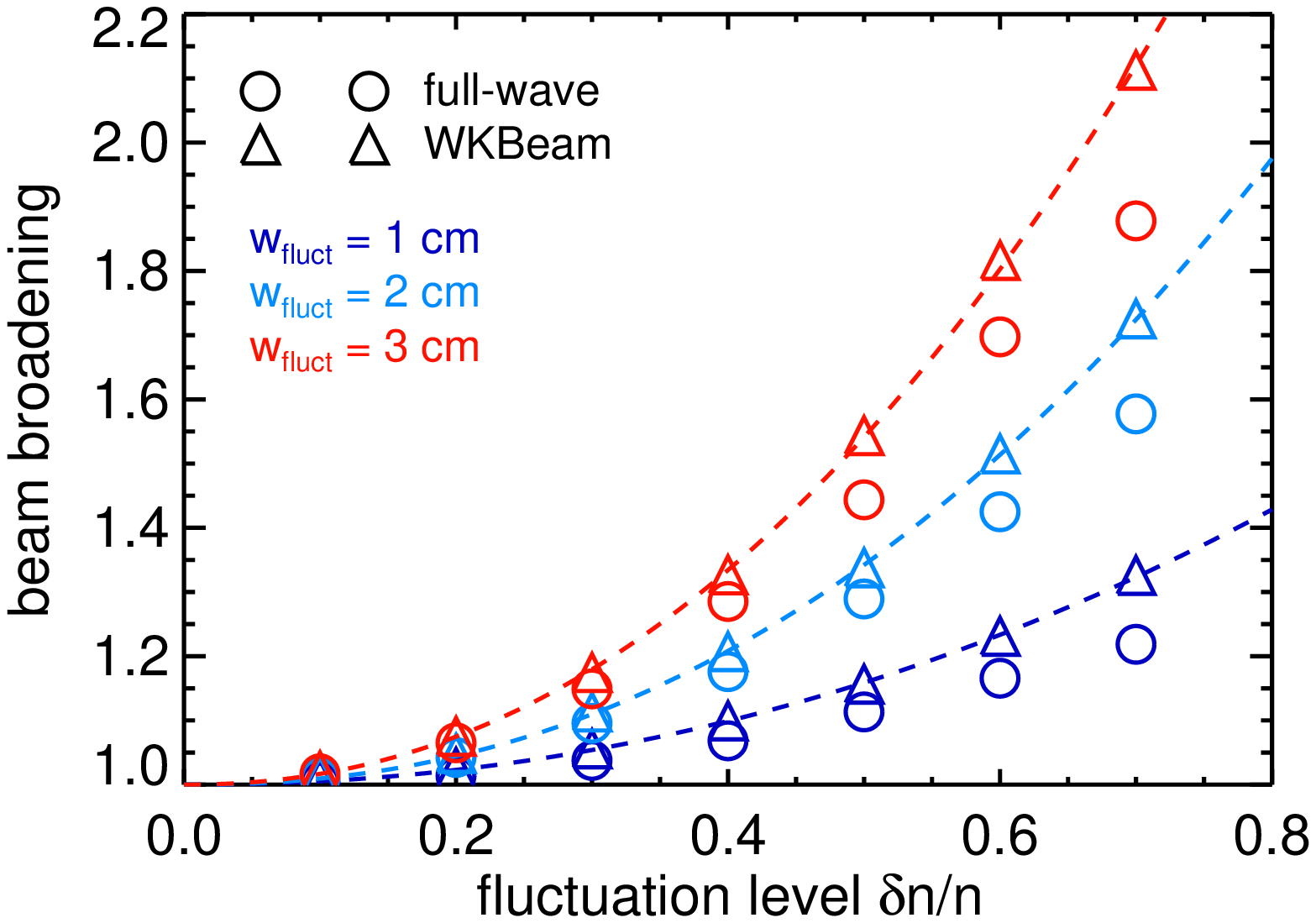}
	\includegraphics[width=0.45\textwidth]{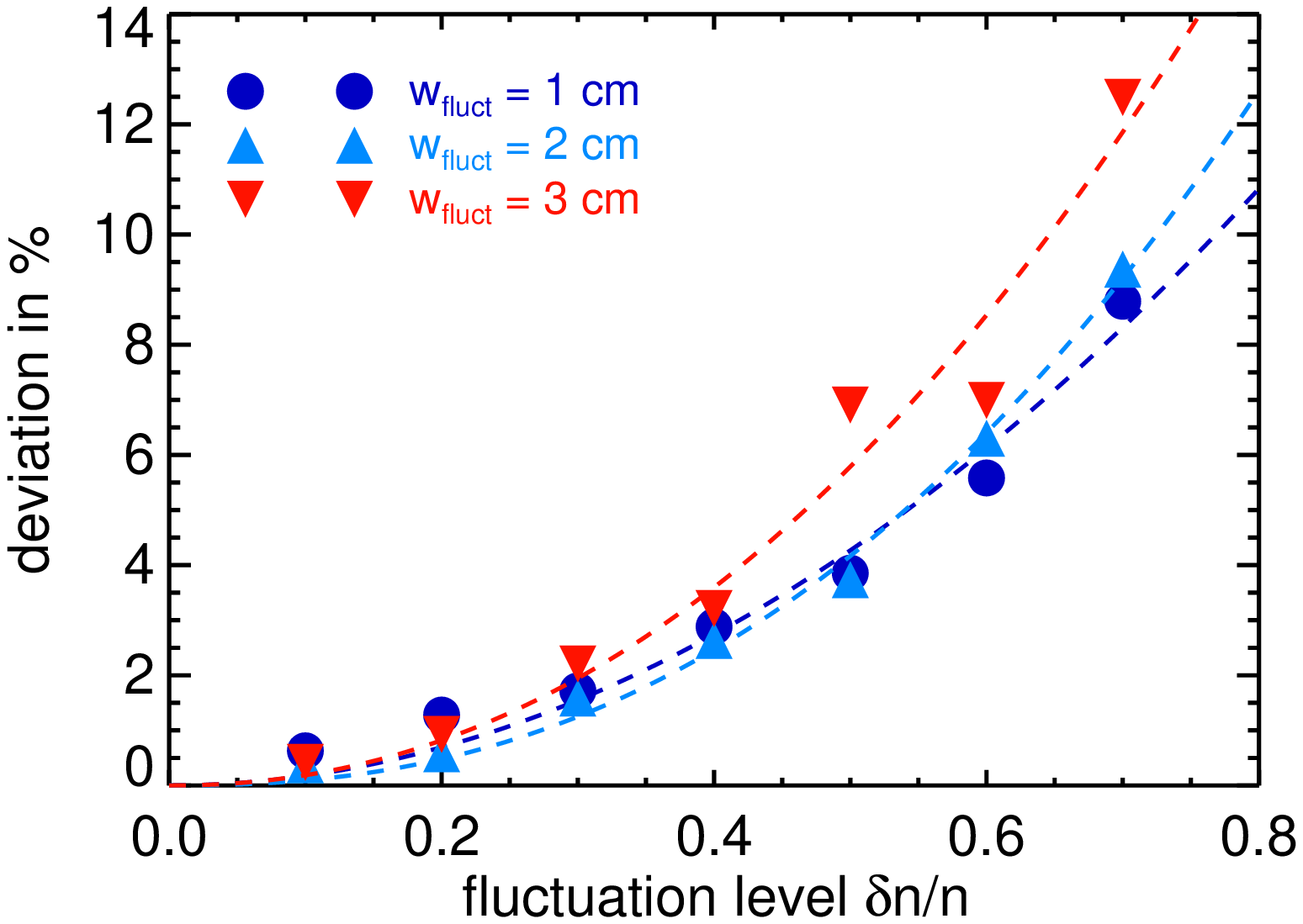}	
	\caption{\emph{(Left)} Beam broadening as a function of the fluctuation amplitude at a detector antenna position of $x=2.31\,\mathrm{m}$ from full-wave simulations and WKBeam calculations as indicated in the plot. Power laws are fitted to the WKBeam results (dashed lines). \emph{(Right)} Percentage deviation of WKBeam to full-wave results using the values shown in the left plot.}
	\label{f:beambroadening_Omode_wfluctScan_gaussw}
\end{figure}

With increasing width of the fluctuation layer the fluctuating density extends to regions of higher density and thus lower $x$ values. Therefore, the position of the detector antennas is, as in the last section, set to a position of $x=2.31\,\mathrm{m}$. 
Figure~\ref{f:beambroadening_Omode_wfluctScan_gaussw} \emph{(left)} shows the beam broadening as obtained from fitting a Gaussian to the detector antenna signals as a function of $A_0$ with $w_{\subtxt{fluct}}$ as parameter. Similar to the results presented so far, the WKBeam calculations yield larger values than the full-wave simulations, with increasing absolute deviation for increasing width of the fluctuation layer. The dependence of the broadening on $A_0$ is found to be stronger for larger value of $w_{\subtxt{fluct}}$, i.e.\ with increasing propagation length in the fluctuation layer, the broadening increases further with maximum beam broadening values of approximately two.

The percentage deviation of the WKBeam results with respect to the full-wave results does not show a strong dependence on $w_{\subtxt{fluct}}$, see Fig.~\ref{f:beambroadening_Omode_wfluctScan_gaussw} \emph{(right)}. The deviations for $w_{\subtxt{fluct}}=1\,\mathrm{cm}$ and $w_{\subtxt{fluct}}=2\,\mathrm{cm}$ as a function of $A_0$ are very similar. Only for $w_{\subtxt{fluct}}=3\,\mathrm{cm}$ a slightly stronger increase with increasing value of $A_0$, i.e.\ a slightly steeper slope, is observed. This is, however, thought to be caused by the increased spatial extension of the fluctuation layer into regions with higher background densities (see Sec.~\ref{s:shift_scan}) instead of an increased propagation distance in the fluctuation layer alone. 

For most of the cases in this scan, the Gaussian provides the better fit to the detector antenna signals, only for large fluctuation amplitude cases ($A_0 \geq 0.5$) at $w_{\subtxt{fluct}} = 3\,\mathrm{cm}$, the Cauchy distribution is the better approximation of the transverse beam profile.

\section{O- and X-mode comparison}\label{s:O_X_mode}
In the cases presented so far, the injected microwave beam was in O-mode polarization. In this section, we present simulations results of WKBeam's capability of injecting a beam in X-mode polarization (by comparing and benchmarking with the corresponding full-wave simulations). Instead of repeating all scans presented so far, from which no further knowledge would be gained, we restrict ourselves to a scan of the fluctuation amplitude $A_0$ for fixed values of $w_{\subtxt{fluct}}=2\,\mathrm{cm}$ and $x_{\subtxt{shift}}=0$. 

\begin{figure}[tb] 
	\centering
	\includegraphics[width=0.45\textwidth]{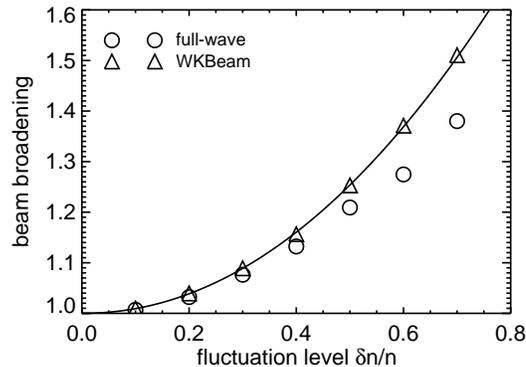}
	\caption{Beam broadening as a function of the fluctuation amplitude at a detector antenna position of $x=2.35\,\mathrm{m}$ from full-wave simulations and WKBeam calculations as indicated in the plot for X-mode injection. A power law is fitted to the WKBeam results.}
	\label{f:OXmode_beambroadening}
\end{figure}


To avoid the right-hand cut-off of the X-mode~\cite{Stix.1992}, we have reduced the background magnetic field to $B_{\subtxt{tor}}=0.25\,\mathrm{T}$ which results in a very similar background density profile when units normalized to the respective cut-offs are considered. Figure~\ref{f:OXmode_beambroadening} shows the beam broadening deduced from the Gaussian fits to the detector antenna signals at a position of $x=2.35\,\mathrm{m}$. The equivalent case for O-mode injection was shown in Fig.~\ref{f:beambroadening_Omode_xshift0_wfluct2} and one can see that they are very similar. For the X-mode case, the power law fitted to the full-wave simulations yields for the normalized beam broadening as a function of the fluctuation amplitude $b=1.05 \cdot A_0^{2.05}$ which is, again, very similar to the O-mode case. The small difference is due to the slightly different normalized background density resulting in slightly different refraction. 


\section{Summary}\label{s:summary}
We have investigated the broadening of a microwave beam passing through a layer of turbulent plasma density fluctuations, resembling the situation of a fusion edge plasma. The results from the WKBeam code were compared with full-wave simulations, performed with IPF-FDMC, over a large parameter range in order to benchmark WKBeam and explore the ranges of validity of the underlying approximations, and specifically the Born approximation. 
\added{This approximation allows to directly calculate the effect of fluctuations in WKBeam by applying a scattering operator whereas the full-wave simulations require an ensemble-average. For the scenarios presented here, this leads to a speed-up of WKBeam of approximately a factor of 4 as compared to the full-wave simulations (scaling the actual wall-clock times of the computations down to a single process). This value will become significantly larger when increasing the wave frequencies which requires a higher spatial resolution and thus larger numerical grids in the full-wave simulations.}

Substantial broadening of the injected microwave beam up to a factor of 2 was found in the scenarios considered. For all cases, WKBeam yielded larger broadening than the full-wave simulations. Up to fluctuation levels of approximately $50\,\%$, however, the overestimation remains below $10\,\%$. If the background density in the fluctuation layer exceeds values of $30-40\,\%$ of the cut-off density of the corresponding mode, the overestimation of WKBeam becomes more pronounced, reaching values of $20\,\%$ at about $70\,\%$ fluctuation level.

The relative deviation from the full-wave solution was found to depend only weakly on the propagation length through the fluctuation layer.

An important observation is the change of the transverse profile of the scattered beam from a Gaussian to a Cauchy distribution for strong scattering. One consequence are the elevated tails of the profile which means more power is scattered into directions far off the original propagation direction.

The parameter range investigated in this paper also includes the ITER scenario recently analyzed with WKBeam: values of $\delta n_e/n_e=20\,\%$ at a normalized background density of $X=0.2$ were assumed~\cite{Snicker.2018}. According to the results presented in this paper, only a small overestimation on the percentage level is expected from WKBeam for these parameters, strengthening the main result from Ref.~\cite{Snicker.2018} that significant beam broadening should be expected for ITER (NTM stabilization should nevertheless still be achievable within the capabilities of the EC upper launcher system). One can furthermore conclude that the interaction of microwaves in the EC range of frequencies with edge density fluctuations can be well described within the limit of the Born approximation in large-scale fusion-relevant tokamak experiments.

No significant difference was found when changing the beam polarization from O- to X-mode. Since both codes are able to investigate X-mode polarized beams, a project to study cross-polarization scattering due to density fluctuations was started~\cite{Guidi.2016}. A thorough benchmark and analysis of this problem is however beyond the scope of this paper and will be published in a following paper.

\section*{Acknowledgments}
This work has been carried out within the framework of the EUROfusion Consortium and has received funding from the Euratom research and training programme $2014-2018$ under grant agreement No.\ 633053. The views and opinions expressed herein do not necessarily reflect those of the European Commission.

Valuable discussions with Dr Gregor Birkenmeier are gratefully acknowledged by one the authors (A.K.). Since most of the paper was written in the train commuting between Munich and Stuttgart, the same author would also like to acknowledge Deutsche Bahn for providing an environment where this is actually possible. The authors are indebted to the efforts of the open-source software community.

\appendix
\section*{Appendix}\label{s:Appendix}
\setcounter{section}{1}

In the full-wave code, the beam waist $w_0$ and the axial distance $x$ to the beam waist are input parameters, whereas in WKBeam, the beam size $w$ and the radius of curvature $R$ in the antenna plane are input parameters. According to e.g.\ Ref.~\cite{Hartfuss.2014}, $w$ and $R$ in the antenna plane are given by the following equations:
\begin{eqnarray}
	w(x) &=& w_0\sqrt{1+\left(\frac{x}{x_R}\right)^2}, \label{e:Gauss_w}\\
	R(x) &=& x\left[ 1+\left(\frac{x_R}{x}\right)^2 \right],\label{e:Gauss_R}
\end{eqnarray}
with $x_R=\pi w_0^2/\lambda_0$ the Rayleigh range. Doing some algebra yields the required expressions for the full-wave code antenna input:
\begin{eqnarray}
	w_0 &=& \frac{w \lambda_0 R}{\sqrt{\lambda_0^2 R^2 + w^4\pi^2}}, \label{e:Gauss_w0}\\
	x &=& \frac{w^4 \pi^2 R}{w^4 \pi^2 + \lambda^2 R^2}, \label{e:Gauss_x}.
\end{eqnarray}
Using the WKBeam input parameters $w = 1.5\,\mathrm{cm}$ and $R = 10\,\mathrm{cm}$, values of $w_0\approx9.7\,\mathrm{mm}$ and $x\approx58.2\,\mathrm{mm}$ are then obtained for the full-wave input parameters\added{.}


\end{document}